# Enhancement of mechanical properties of high modulus polypropylene grade for multilayer sewage pipes applications

*Helena KHOURY MOUSSA\*, Georges CHALLITA, Houssem BADREDDINE, Guillaume MONTAY, Bruno GUELORGET, Thomas VALLON, Wadih YARED, Marwan ABI RIZK, Akram ALHUSSEIN*

H. Khoury Moussa, H. Badreddine, G. Montay, B. Guelorget, T. Vallon

UR LASMIS, Université de Technologie de Troyes, 12 Rue Marie Curie, CS 42060, 10004 Troyes, France.

E-mail : Helena.khoury_moussa@utt.fr

H. Khoury Moussa, G. Challita

Équipe MMC, CRSI, Lebanese University, Faculty of Engineering, Roumieh, El-Metn, Lebanon.

W. Yared, M. Abi Rizk

Advanced Plastic Industries – API Company, Dbayeh highway, Seaside road, Dbayeh, El-Metn, Lebanon.

A. Alhussein

UR LASMIS, Université de Technologie de Troyes, Pôle Technologique Sud Champagne, 26 rue Lavoisier, 52800 Nogent, France.

## Abstract

Advances in technology have provided fresh generations of stiff polypropylene block copolymers for gravity sewerage applications. The aim of this study is to further enhance the stiffness of these materials through the incorporation of inorganic fillers. In this study, three talc filled PP and one glass fiber filled PP composites were characterized in order to be used as a middle layer in a three-layer sewage pipe. The obtained results showed an increase of approximately more than 100 % and 250 % in tensile and flexural moduli by the use of 30 - 50 wt.% talc-filled PP and 30 wt.% glass fiber-filled PP, respectively. This high increase in the rigidity of the material would allow manufacturing pipes with improving ring stiffness. Composites filled with 30 wt.% talc or glass fiber showed good filler-matrix interaction and good filler distribution and dispersion. However, reduced filler-matrix interaction was observed in the case of the composite filled with 50 wt.% talc. In addition, the use of Differential Scanning Calorimetry analysis revealed that the addition of fillers enhanced the crystallization temperature of the polypropylene matrix. Furthermore, Thermogravimetric Analysis showed that the high modulus PP grade retained its thermal stability in the various composites.

**1. Introduction**

Majority of old sewer pipes have been made from brittle rigid materials such as concrete or clay, which fail due to stone impingement or ground movement. Internal corrosion of concrete sewer pipe is caused by sulfuric acid generated by domestic/municipal wastewater, which further weakens the structure and is a severe concern in hot dry climates. [1–3] Traditional sewage systems are also plagued by ground movement and tree root intrusion. [4] Plastics pipeline systems have been introduced during the early days of plastics, and since, have been refined and developed to become a major material of choice for gas, water, sewage, and industrial pipelines. Flexible plastic pipe systems are now able to meet the majority of the requirements of modern sewage pipe systems. They also have additional advantages over rigid materials, such as; no brittle pipe failures due to overloading, low pipe weight, lower risk of leakage at the connection points, excellent chemical resistance, long lifetime, an environmentally friendly solution of the installed pipe systems, possibility to bend the plastic pipe to be installed in curves, relatively high pipe length, and easy jointing. [5–10] A Finite Element Model (FEM) of a Polyethylene (PE) pipe loaded by internal pressure, residual stresses and soil loads was developed by Poduska et al. [11] According to their study, the lifetime of PE pipes exceeds 100 years. Polyvinyl chloride (PVC), High Density Polyethylene (HDPE), and Glass Fiber Reinforced Plastic (GFRP) were widely employed in sewage applications. However, Polypropylene (PP) has gradually acquired a greater place over the last few decades as a pipe material for non-pressure sewerage applications, this is due to its superior stiffness to weight ratio, high temperature resistance, chemical resistance across a broad pH range, good abrasion resistance, and excellent long-term properties. [12] Advances in technology have provided fresh generations of stiff PP block copolymers for gravity sewerage applications, resulting in very stiff PP block copolymers with high-impact and excellent long-term properties. High Modulus

Polypropylene (PP-HM) grades (1700 MPa and higher E-modulus) are stiffer compared to alternative materials such as HDPE (typical E-Modulus of 1100 MPa). Typically, PP-HM grades are reactor-made products consisting of an extremely crystalline PP-H matrix that contributes to the stiffness, in which an ethylene/rubber phase is finely dispersed, contributing to elevated impact resistance. [13]

Incorporation of inorganic particle fillers has been found to be an effective way to improve mechanical properties of a PP matrix and / or to reduce the costs. Reinforcements, which are much stiffer and stronger than the polymer, generally improve its modulus and strength. [14] Consequently, the modification of mechanical properties can be regarded as their main function. The effects of filler loading on the mechanical, flow, and thermal properties of PP composites filled with talc, kaolin, and Calcium Carbonate ($CaCO_3$) were investigated by Leong et al. [15] They reported that talc and kaolin were found to reinforce PP by increasing tensile and flexural moduli with filler load. However, at higher filler loading, talc tends to agglomerate, resulting in a significant loss in the strength and toughness of talc-filled PP composite. $CaCO_3$-filled PP composites have very high resistance to impact. However, the rise in PP – $CaCO_3$ composites impact toughness is accompanied by a decrease in tensile and flexural strength. Lapcik et al. [16] found that increasing talc content led to an increase in the mechanical strength of the talc–PP composite material with a simultaneous decrease in the fracture toughness. Also, with increasing talc content, the thermooxidative stability increased. Furthermore, it was proved that with increasing talc content a significant increase in Young's modulus [17,18], indentation modulus, and hardness [19] was observed. Another finding [20] reported that the stiffness of talc filled PP composites increases as the talc loading increases and decreases as the test temperature increases. However, the yield strength decreases as the temperature and talc content increase. Wang et al. [21] demonstrated that the addition of talc fillers to the PP matrix led to an increase of the thermal stability, melting temperature, crystallization temperature, crystallinity percentage, and Young's modulus, in addition to a decrease of the glass transition temperature and the elongation behavior. Moreover, they observed an increase of the yield strength at low tensile velocity and a decrease of it at high tensile velocity. The mechanical and thermal properties of hybrid PP composites and single-filler PP composites were studied by Leong et al. [18] They found that the tensile and flexural strength and modulus of talc-dominant hybrid composites were higher, while $CaCO_3$-dominant hybrids were tougher and more deformable. In an investigation [22], it has been reported that increasing the appropriate percentage (by weight) of talc improved mechanical properties. In terms of stiffness, modulus, and heat deflection temperature, 30–50% talc filled PP show the best results. However, tensile strength decreases as the percentage of filler increases. To increase the inherent properties of modified PP, comparative studies using talc and Moroccan clay were conducted. [23] With particle loading ranging from 10% to 35%, an increase in thermal degradation temperature of all composites was observed. In addition, the crystallization temperature of the composites increased somewhat as the filler concentration in the polypropylene increased with no discernible effect on the melting temperature of the PP matrix. The rise in crystallinity of the filled polypropylene composites was also related to the improvement in tensile properties of talc and clay filled polypropylene. Tensile strength decreased linearly with increasing particle content in all composites. All composites showed a progressive decrease in strain at yield as particle loading increased from 10% to 35%. Also, talc is a strong nucleating agent; talc addition causes a rising of the PP crystallization temperature and crystallinity. [15,20,21,23–25] The composite is expected to have a greater modulus, better dimensional stability, and improved strength as the crystallinity of the polymer matrix increases, but a lower elongation at failure. This nucleating effect is slightly stronger than for standard talc having a surface area around three times lower than that of submicronic-talc. [26] However, the effect of crystallinity on the mechanical properties of $CaCO_3$ and kaolin-filled PP composites is modest because both of these fillers have weak nucleating abilities when compared to talc. [15]

The influence of temperature, moisture, and hygrothermal aging on the tensile behavior of thermoplastic composites was investigated experimentally. [27] Large deformations were observed for talc-filled PP (PP-T) and short glass fiber filled PP (PP-G) at 85 °C and 120 °C, respectively. At all temperatures, the plastic deformation of PP-G was smaller than that of PP-T, however, the strength values of PP-G were higher than PP-T, which was due to the strengthening effect of glass fibers. Tensile properties decrease linearly with temperature. The tensile strength, elastic modulus (mostly for PP-T), and strain at failure of PP-T and PP-G were all affected by hygrothermal aging. Because talc decomposes at a considerably higher temperature than PP, the addition of talc filler raised the degradation temperature of PP/talc composites. [21] It was also found by Zhou et al. [28] that incorporating talc particles into a polypropylene matrix can enhance modulus and decomposition temperature while decreasing yield strength and fatigue strength and having no effect on glass transition temperature and melt temperature. Panthapulakkal and Sain [29] studied the mechanical, water absorption and thermal properties of injection-molded Short Hemp Fiber/Glass Fiber-filled PP hybrid composites. They found that glass fiber incorporation improves the tensile, flexural, and impact properties of short hemp fiber composites. In another study, [30] 30 wt.% glass fibers, talc or $CaCO_3$ were introduced into a recycled PP (RPP) matrix in order to obtain properties close or greater than those of the original PP. The three types of fillers increased the elasticity modulus of the RPP polymer. However, tensile, yield, and impact strengths were decreased with the addition of talc and $CaCO_3$, while a sharp increase was obtained by the use of glass fiber. Also, a better filler matrix interface was observed for glass fiber filled RPP.

The combination of individual layers, in multilayer pipe systems, allows all the distinct requirements imposed on sewage pipes to be achieved. Multilayer pipes are used wherever the benefits of thermoplastics are desirable and high rigidity is also needed, such as non-pressure sewage pipes. Hutar et al. [31] used a fracture mechanics approach to assess the failure of a three-layer composite plastic pipe consisting of two protective layers and a main HDPE pipe layer. They reported that, under certain conditions, a crack can be stopped at the interface, significantly increasing the pipe lifetime. Farshad [32] demonstrated a new methodology for predicting the long-term behavior of multilayer plastic pipes with metallic interlayers, single layer pipes, and structured pipes. This methodology, helped identifying the first layer failing by creep under internal pressure and quantifying the additional capacity of the remaining layer to hold the hydrostatic pressure. Furthermore, Nezbedova et al. [33] proposed a new method for estimation of the lifetime of a PE multilayer pipe by measuring basic fracture parameters and calculating the corresponding fracture toughness values of PE pipes using FEM. Moreover, Arbeiter et al. [34] presented results from cyclic fatigue tests on un-reinforced PP block copolymer, as well as crack growth kinetics for the mineral-reinforced middle layer at various loading ratios. Slow crack growth was observed in both reinforced and unreinforced polypropylene block copolymer pipes. On the other hand, Estrada et al. [35] used the finite element method to investigate the effect of the boundary condition on the first ply failure and stress distribution of a multilayer composite pipe.

Ring stiffness is the main property that characterizes a non-pressure pipe. It refers to the pipe ability to withstand ring deflection caused by radial forces. The ring stiffness of a pipe depends on the stiffness of its material and the structural design of its wall. The ring stiffness test was carried out by Seibi et al. [36] on four different shapes of glass-reinforced plastic pipes: box, flat elliptic, egg, and round. The stiffest and most flexible pipe shapes among all tested pipes are circular and flat elliptical pipes, respectively. Park et al. [37] investigated the pipe stiffness of GFRP flexible pipes using the parallel plate loading test, FEM, and theoretical analysis. Furthermore, many researchers have published a lot of information on the stiffness of flexible plain pipes. [8,37–40] In order to calculate the elasticity modulus, Thornblom et al. [41] used the parallel plate test on four different PP grades. This latter was found to be proportional to the ring stiffness. In addition, Wierzbicki and Szymiczek [42] used the parallel plate device in order to measure the ring stiffness of a multilayer PP pipe.

Multilayer composite pipes with a high ring stiffness are now being produced by various sewage pipe manufacturers. These pipes have a number of advantages, including strong impact and abrasion resistance, high chemical resistance, and thermal stress resistance, a smooth pipe inner surface, and a large range of appropriate fittings. However, there is still no agreement on the optimal configuration for such a pipe. In a previous study, [43] a FEM was developed to predict the ring stiffness of various multilayer pipe configurations as a function of talc content in the composite layer and layer thicknesses.

Commercially, Borealis introduced products with flexural / tensile E-modulus values up to 1800 MPa, [44,45] representing a 20-50 percent comparative rise over conventional PP block copolymer products used for sewerage applications with flexural modulus / tensile E-modulus values of approximately 1200 to 1300 MPa. The fresh generation of PP-HM materials with enhanced E-Modulus offers greater resistance to creep and stress relaxation. These materials have been specifically produced for sewage pipe systems and demonstrate great stress cracking resistance. The aim of this study is to further enhance the stiffness of these materials through the incorporation of inorganic fillers, permitting for thinner pipe walls, reduced pipe weight, and important cost savings as a result. It is predicted to use these materials as a middle layer in a three-layer sewage pipe. This research is being conducted in collaboration with Advanced Plastic Industries (API, Lebanon, Zouk Mosbeh, www.api.com.lb), a company that manufactures plastic pipes and fittings. The company is investing in this research to develop a new sewer pipe with improved mechanical properties and a lower cost while retaining the ideal chemical resistance of PP pipes.

## 2. Experimental Procedure

### 2.1. Materials

A high modulus PP block copolymer **BA212E** (BorECO BA212E – Borouge, Borealis Group, 1 George Street #18-01 Singapore 049145) with a melt flow index equal to 0.3 g/10min, a PP homopolymer **PPH2250** (TASNEE PPH2250 – TASNEE, 11496 Kingdom of Saudi Arabia) with a melt flow index equal 25 g/10min, a talc grade **Mikron's talk premier** (Mircon'S, 34674 Üsküdar, Turkey) with a mean particle size of 5.3 μm, and glass fiber **DS 1125-10N** (BRAJ BINANI Group, B-1560 Hoeilaart, Belgium) with 10 μm fiber diameter and 4 mm fiber length, were used in this study.

### 2.2. Preparation of Talc/PP and Glass Fiber/PP Composites

Talc and glass fiber filled PP polymer composites were produced using CHT20 (Qingdao, China) twin screw extruder. The extruder temperature was in the 160-230 °C range for filled PP. [15,18,30] Four different composite materials were produced; PP-MT30 (A) and PP-MT30 (B) are 30 wt.% talc-filled PP composites with different matrix composition, PP-MT50 is 50 wt.% talc-filled PP, and PP-GF30 is 30 wt.% glass fiber-filled PP. **Table 1** lists the different materials and their corresponding composition. Composites were extruded, cooled and then granulated.

**Table 1**. Composition of the prepared composite materials.

| Material | Component (wt.%) | | | |
|---|---|---|---|---|
| | BA212E | PPH2250 | Talc | Fiber Glass |
| PP-MT30 (A) | 50 | 20 | 30 | 0 |
| PP-MT30 (B) | 55 | 15 | 30 | 0 |
| PP-MT50 | 25 | 25 | 50 | 0 |
| PP-GF30 | 50 | 20 | 0 | 30 |

## 2.3. Melt Flow Index

Melt Flow Index (MFI) is defined as the mass of polymer, in grams, flowing in ten minutes through a capillary of a specific diameter and length at a defined temperature T and pressure. MFI is measured by a melt flow indexer (SCITEQ, MFI 450 series, Sciteq Denmark), based on the standard ISO 1133 [46] at T = 230 °C with a nominal load $M_{nom}$ equal to 2.16 kg and a cut-off time interval t of 120 seconds, according to *Equation* **1.**

$$MFI(T, M_{nom}) = \frac{600 \times m}{t} \quad [g/10\,min] \tag{1}$$

*Where, m* is the average mass of the cut-offs, in grams.

## 2.4. Density Determination

The density (ρ) is measured based on the ISO 1183 [47] immersion method (method A) using a Mettler Toledo balance (Mettler Toledo, ME204, Technoline), according to the following equation:

$$\rho = \frac{m_{S,A} \times \rho_{IL}}{m_{S,A} - m_{S,IL}} \tag{2}$$

Where,

$m_{S,A}$ is the apparent mass, in grams, of the specimen in air;

$m_{S,IL}$ is the apparent mass, in grams, of the specimen in the immersion liquid (water);

$\rho_{IL}$ is the density of the immersion liquid (water) at 23 °C, in grams per cubic centimeter.

**2.5. Mechanical Testing**

Specimens for different tests were produced at API by injection molding. Test specimens were injection-molded in a tensile mold with a BMB Mc150 (B.M.B, 6901181, B.M.B Italy) machine. The following conditions were used during the injection-molding phase: screw speed 100 rpm, cooling time 15 s, injection pressure 80 bar, the injection-molding temperature ranged from 200 to 225°C, displayed on five different zones along the screw.

*2.5.1. Tensile Properties*

Tensile tests were conducted on the different materials according to ASTM D638. [48] The test method depicted in ASTM D638 covers the determination of the tensile properties of unreinforced and reinforced plastics in the form of standard dumbbell-shaped test specimens, as illustrated in **Figure 1**. All reported results present the averages of five measurements for each material.

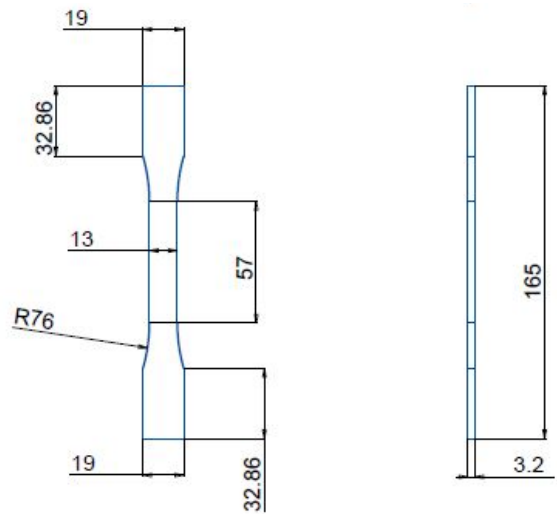

**Figure 1.** Tensile test specimen (dimensions in mm).

Specimens were mounted and loaded on a tensile machine (INSTRON-4411 - 5kN, England). The specimen was loaded up to failure with a loading velocity of 5 mm/min at room temperature. The local displacements and strains in the useful part were measured by a video extensometer which is based on Digital Image Correlation (DIC) technique. The video extensometer system consists of two high resolution cameras, and a DIC post-processing software (Davis, LaVision GmbH, Göttingen, Germany) as illustrated in **Figure 2**. According to Quanjin et al., [49] 50 DIC method offers reasonable results compared to the experimental and strain gauge measurements methods. Black speckled pattern was created using an ink pad (Correlated solutions, .013'' dot size, Archival ink) at the outer surface of the gauge region of the specimen as shown in **Figure 3 (a)** for all specimens except for PP-GF30 the outer surface of the gauge region of the specimen was sprayed with a white color (White gloss Spray Paint, RS 764-3023, RS PRO) to create an irregular speckled pattern as shown in **Figure 3 (b)**.

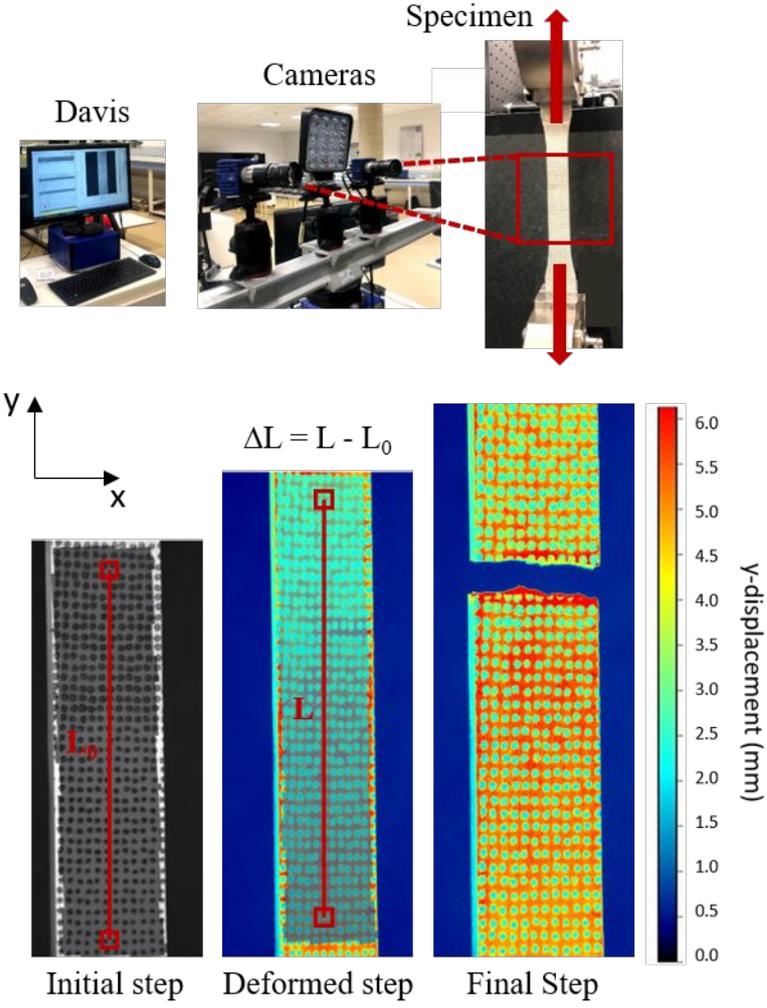

**Figure 2.** Digital image correlation setup.

After preparing the specimen with the adequate pattern, the measurement setup, and the loading device, images before and after deformation were recorded. Then, the displacements and the strains were obtained; the movement of the speckles was tracked and correlated to the original undeformed pattern.

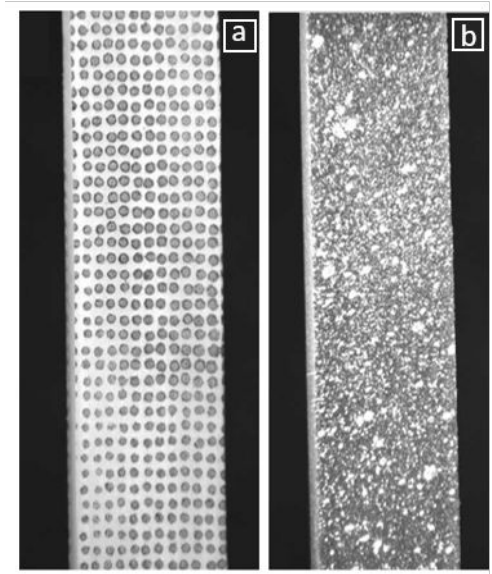

**Figure 3**. **(a)** Black speckled pattern and **(b)** irregular white speckled pattern.

*2.5.2. Flexural Properties*

Flexural tests were conducted on the different materials according to ASTM D790-03. [51] The test method depicted in ASTM D790-03 covers the determination of flexural properties of unreinforced and reinforced plastics, including high-modulus composites in the form of rectangular bars, as illustrated in **Figure 4 (a)**. All reported results present the averages of five measurements for each material.

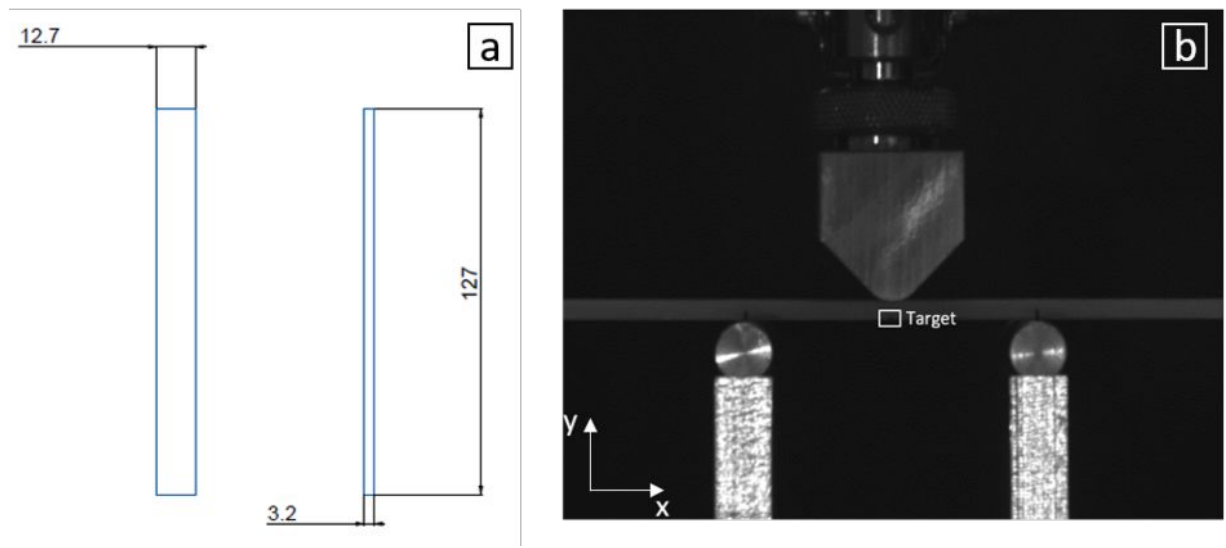

**Figure 4.** Flexural test **(a)** specimen (dimensions in mm) and **(b)** setup.

A three-point bending test was performed where the specimen was installed on two supports separated by 50 mm, and loaded by means of a loading nose midway between the supports. The specimen was loaded up with a loading speed of 2 mm min$^{-1}$ at room temperature. The load was registered using a data acquisition system. To accurately measure the flexural modulus of the materials, strain was calculated by following the y-displacement of a target point installed on the lower surface of the specimen under the loading nose, as illustrated in **Figure 4 (b)**, using a camera and the software Fringe Analysis (FA4, Holo3, St-Louis, France).

Flexural stress is calculated for any point on the load-deflection curve by means of the thin beam model giving the following equation:

$$\sigma_{xx} = \frac{3PL}{2bd^2} \qquad (3)$$

where:

$\sigma_{xx}$ = stress in the outer fibers at midpoint, in MPa,

P = load at a given point on the load-deflection curve, in N,

L = support span, in mm,

b = width of beam tested, in mm,

d = thickness of beam tested, in mm.

Flexural strain is calculated for any deflection using **Equation 4**:

$$\varepsilon_{yy} = \frac{6 U_D d}{L^2} \qquad (4)$$

where:

$\varepsilon_{yy}$ = strain in the outer surface, in mm mm$^{-1}$,

$U_D$ = maximum deflection of the center of the beam, in mm.

## 2.6. Thermal Testing

*2.6.1. Differential Scanning Calorimetry*

The crystallization behavior and melting characteristics of the materials were studied by Differential Scanning Calorimetry (DSC) according to ISO 11357-3. [52] Measurements were carried out on a simultaneous TGA-DTA/DSC SETARAM instrument (Caluire, France), with an argon circulation at a heating and cooling rate of 10°C/min. In the first heating and cooling scans, the samples were heated from 25°C to 220°C and held at that temperature for 5 min to eliminate the thermal history; then, the nonisothermal crystallization process was recorded from 220 to 25°C, and a standard status of crystallization was created; the crystallization temperature ($T_c$) of PP phase was measured from this cooling stage. Concerning the melting temperature of PP ($T_m$), it was measured during a second heating stage performed with the same conditions as the first one.

*2.6.2. Thermogravimetric Analysis*

The thermal stability of the materials as well as the filler content in the composite materials were investigated by Thermogravimetric Analysis (TGA) using a simultaneous TGA-DTA/DSC SETARAM instrument (Caluire, France) according to ASTM E1131-08. [53] Samples were introduced in alumina crucibles and were heated up to 600˚C under argon atmosphere, at 600˚C the purge gas flowing over the sample was automatically switched to oxygen and the sample was further heated to 750 ˚C, with a heating rate of 10 ˚C/min.

## 2.7. Observation of Materials Morphology

The mode of fracture and the morphologies of talc and glass fiber reinforced polymer composites were analyzed by observing the fracture surfaces of tensile samples. This was carried out with a scanning electron microscope (SU8030, Hitachi); the fracture surfaces of the tensile specimens were sputter-coated with a thin gold–palladium layer to prevent electrical charge accumulation during the examination and analyzed at an accelerating voltage of 10 kV.

## 3. Results and Discussion
## 3.1. Melt Flow Index

The MFI values of BA212E and the various composites are shown in **Figure 5**. The high stiffness of BA212E compared to other PP grades is due to its low MFI; 0.33 g/10min. In fact, the low MFI value indicates higher molecular weight, [54] which results in enhanced stiffness. It is also known that the incorporation of fillers into a polymer matrix hinders the plastic flow and increases the viscosity of the polymer melt. Therefore, MFI is expected to decrease with filler loading. As BA212E possesses a very low MFI value, adding fillers to it will result in an unprocessable material. On that account, BA212E was blended with PPH2250 which is a polypropylene homopolymer with a melt flow rate equal to 25 g/10min. [55]

It has been found for a three-layer pipe that, in particular, if the MFI value of the core layer is considerably higher than the MFI values of the inner layer and the outer layer, many problems arise during extrusion. [56] In particular, the flow through the extrusion die is problematic; the core material with the lower viscosity readily flows through the extrusion die, whereas the material for the inner and outer layers does not. The latter layers are not carried along by the core as well as desired, resulting in an irregular thickness of the inner and/or outer layer. Constituently, the composite material with the lower viscosity readily flows through the extrusion die, covering the whole thickness of the pipe. The PP layers are not carried along by the composite layer as well as desired, resulting in an irregular thickness of the PP layers. To solve such a problem, it is recommended that the MFI values and therefore the viscosities of all the plastics should be equal under the conditions of processing. The processing conditions selected for each of the layers are selected in such a manner that the viscosities or MFI values of all the layers are substantially equal. If it is attempted to produce multilayer pipes in a conventional way, materials of approximately equal MFI values must be chosen for all layers in order to achieve a good product.

Accordingly, the first goal of this work was to produce composite materials having MFI values close to that of BA212E, which will be used for the inner and outer layers. **Figure 5** indicates that PP-MT30 (B) and PP-GF30 are the best candidates.

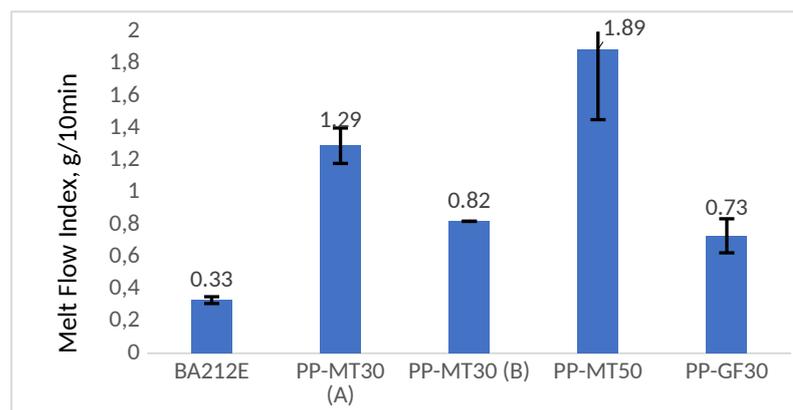

**Figure 5.** MFI values of BA212E and the composite materials.

## 3.2. Mechanical Properties

### 3.2.1. Tensile Properties

**Figure 6** shows the y-displacement maps for the six materials and the corresponding points on the tensile engineering stress – strain curves. Points 1 till 5 correspond respectively to 10, 20, 50, 80, and 100% of the tensile strength (except for PP-MT30 (A) point 5 corresponds to 87% of the tensile strength). This selection of stresses ratio allows the comparison of specimens with the same load carrying capability and durability, preventing data misinterpretation if the comparison is made under the same stress value.

For all the materials it is clearly seen from both the contour plots and the curves that points 1, 2, and 3 are in the elastic zone and thus could be used to calculate the elastic modulus. Indeed, the displacement maps for these three points show a homogeneous displacement field. In addition, these points are in the linear part of the stress-strain curves.

Higher displacement values are noticed from point 4 which correspond to approximately the beginning of the plastic deformation region. When comparing figures that correspond to 100% of tensile strength, one could remark that higher displacement levels, thus higher strain values are obtained for BA212E and PPH2250; they have a more heterogeneous displacement field throughout the gauge length. However, composite materials showed a more uniform strain distribution. The patchy appearance of strain field suggests that pure PP materials are not transferring the load competently to gauge length of tensile sample as compared to composites, thus resulting in increase of both axial and transverse strain gradients. We can deduce from this observation that composites materials are able to withstand higher stress levels.

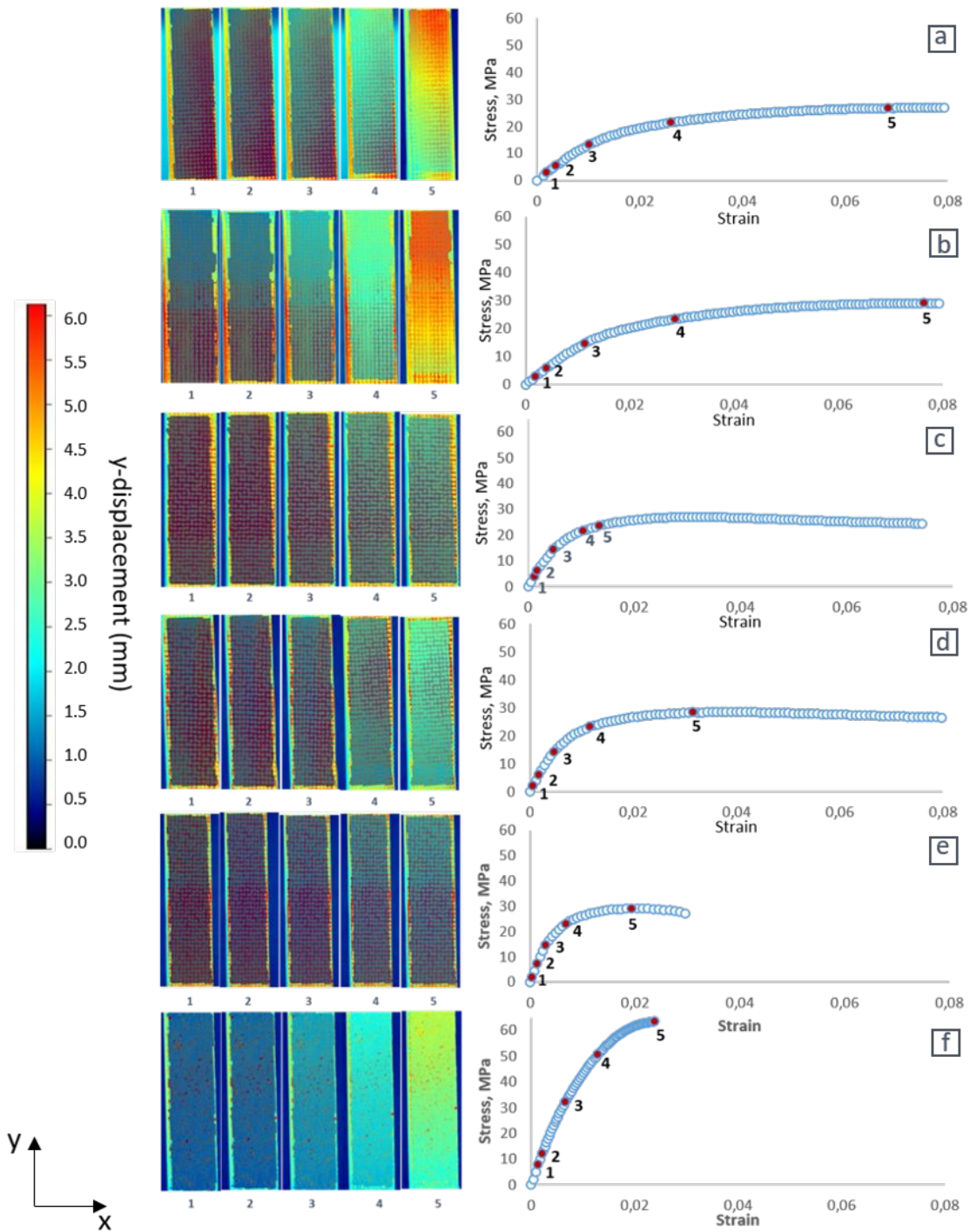

**Figure 6.** Y-displacement field maps and tensile stress-strain curves for **(a)** BA212E, **(b)** PPH2250, **(c)** PP-MT30 (A), **(d)** PP-MT30 (B), **(e)** PP-MT50 and **(f)** PP-GF30.

Filling a polypropylene matrix with talc or glass fiber will result in enhanced mechanical properties. Tensile strength values are shown in **Figure 7**, 30 wt.% and 50 wt.% talc filled-PP did not affect the tensile strength of BA212E. However, 30 wt.% GF-filled PP increased the tensile strength of BA212E by 126 %. Similar results were obtained by Panthapulakkal and Sain. [29] This increase in the strength is expected since glass fiber is stronger and stiffer than talc.

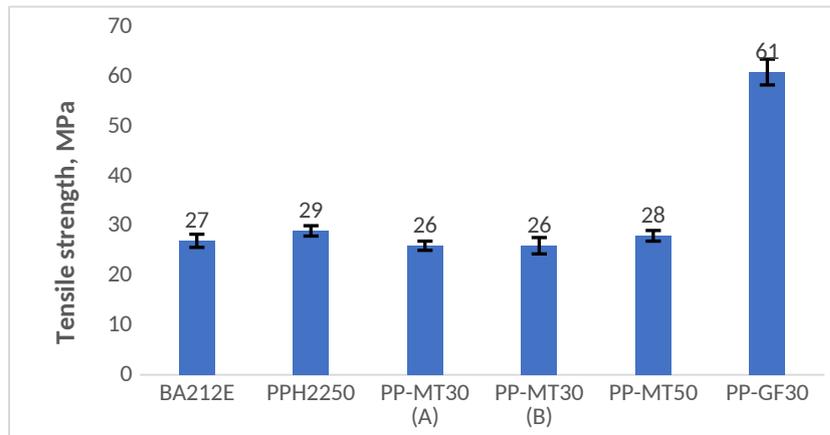

**Figure 7.** Tensile strength of the different materials.

The effect of talc and glass fiber addition on the tensile modulus of BA212E is shown in **Figure 8**. The elasticity modulus increased with the addition of talc and glass fiber into the PP matrix. The 1416 MPa elastic modulus of BA212E was increased by approximately 126% by the addition of 30 wt.% talc. Larger improvement was observed for PP-MT50 composite where the modulus of BA212E reached 5296 MPa which correspond to an increase of 274%. In fact, many researchers reported that increasing filler loading will result in increased moduli. On the other hand, 30 wt.% glass fiber filled PP enhanced the modulus of BA212E by 288%. This result is consistent with previous studies. [30,57,58]

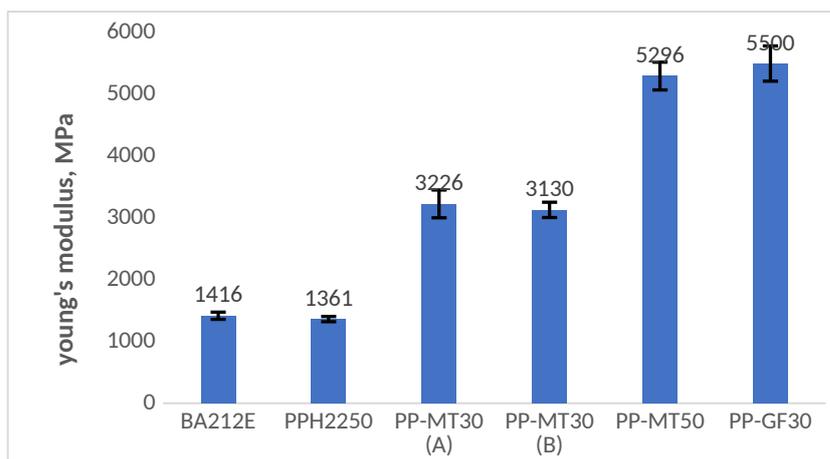

**Figure 8.** Tensile modulus of the different materials.

*3.2.2. Flexural Properties*

Flexural strength and modulus of the different materials are shown in **Figure 9** and **Figure 10**, respectively. Flexural strength of BA212E was found 37 MPa and it was improved significantly by the addition of 30 wt.% glass fiber leading to an improvement of 149%.

However, the PP filled with 30 wt.% or 50 wt.% talc did not show any noticeable increase in its flexural strength.

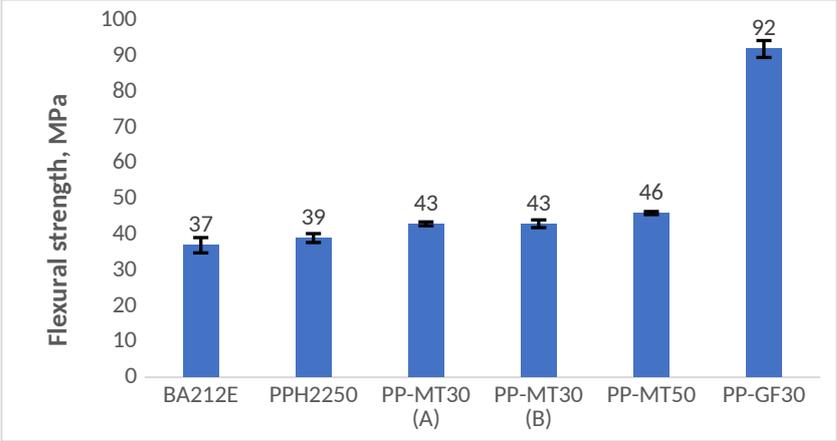

**Figure 9.** Flexural strength of the different materials.

On the other hand, adding 30 wt.% talc to the polymer blend improved the flexural modulus of BA212E by more than 100%. Increasing the talc content from 30 to 50 wt. % enhanced the flexural modulus from approximately 3100 to 4810 MPa. For glass fiber filled PP the modulus of BA212E was improved by 250%.

Despite the fact that the failure in three-point bending tests is due to bending and shear failure, the enhancement in flexural strength and modulus is consistent with the tensile strength and modulus results.

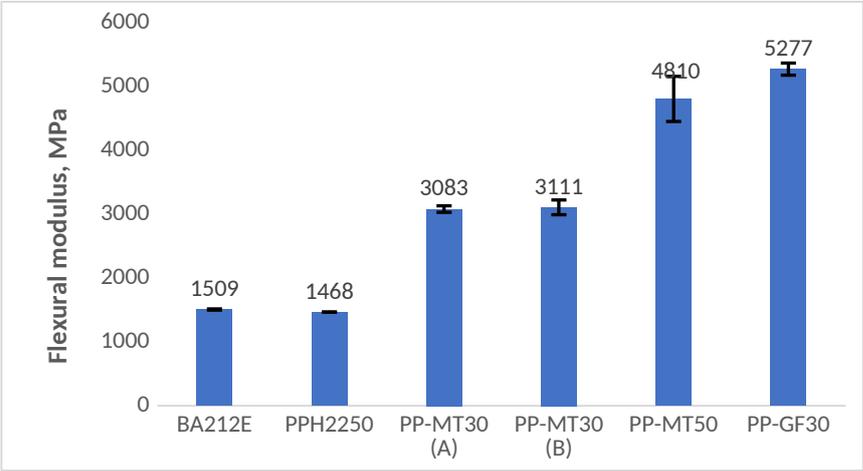

**Figure 10.** Flexural modulus of the different materials.

In both flexural and tensile tests, glass fiber showed better improvement than talc, and this may be due to the fact that talc tends to agglomerate because of its fine particles, as a result, the polymer matrix mobility is increased, and the composites overall stiffness is reduced [15].

The flexural modulus of composites is often lower than the tensile modulus. In fact, the surfaces of composites have an excess of polymer due to the constraints imposed by mold walls.

Thereby, in flexural testing where the maximum stress is at the surface, the surface properties are emphasized at the expense of the interior, resulting in lower measured modulus values.

The addition of talc or glass fiber significantly increases the modulus of the matrix, which is the target parameter of this study. As the ring stiffness is directly related to the stiffness of the material; a pipe made only of BA212E that has a tensile modulus equal to 1416 MPa, will have lower ring stiffness than a multi-layer pipe that has a composite middle layer with a modulus at least 100% greater than that of BA212E. Thus, multilayer pipes can exhibit higher ring stiffness than plain pipes of the same thickness. It is therefore obvious that the use of multilayer pipes will allow the production of a sewer pipe with a reduced wall thickness while maintaining the same ring stiffness value as that of a plain PP pipe.

### 3.3. Thermal and Physical Properties
*3.3.1. Thermogravimetric Analysis Results*

Density values as well as results obtained from TGA are listed in **Table 2**. The onset decomposition temperature ($T_{onset}$) of BA212E is higher than that of PPH2250, while those of the different composites are close to that of BA212E. Thus, the developed composites conserve the excellent thermal stability of BA212E. Furthermore, $T_{1/2}$ (temperature corresponding to 50% weight loss) values for the two PP grades and the talc-filled composites are very close. However, $T_{1/2}$ of PP-GF30 is approximately 10°C higher. Because BA212E and PPH2250 are pure PP and decompose well before 750 °C, the residue remaining after this temperature is negligible. However, in the case of composite materials, this residue serves as the actual reinforcement content. It is evident that the density of the matrix increases after the reinforcement is added. PP-MT50 showed the greatest increase followed by PP-MT30 (A). The density of pure PP increases by 30% and 45% after adding PP-MT30 (A) and PP-MT50, respectively. For PP-MT30 (B) and PP-GF30, the increase was 22% and 26%, respectively. Since these composite materials with enhanced mechanical properties will be used as a core layer in a three-layer pipe with a wall thickness much lower than a plane PP pipe, this increase in density will not increase the weight of the pipe. In fact, a much smaller amount of material will be used.

**Table 2**. Density and thermal properties of the different materials.

|            | $T_{onset}$ (°C) | $T_{1/2}$ (°C) | Residue remaining after 750°C (%) | Density (g/cm³) |
|------------|------------------|----------------|-----------------------------------|-----------------|
| BA212E     | 438.4            | 461.9          | 2.29                              | 0.904           |
| PPH2250    | 431.1            | 459.3          | 3.10                              | 0.904           |
| PP-MT30 (A)| 437.1            | 458.5          | 29.01                             | 1.172           |
| PP-MT30 (B)| 440.6            | 461.8          | 26.90                             | 1.101           |
| PP-MT50    | 436.7            | 454.9          | 44.60                             | 1.308           |
| PP-GF30    | 435.5            | 471.7          | 32.31                             | 1.136           |

*3.3.2. Differential Scanning Calorimetry Results*

**Table 3** summarizes the results of melting and crystallization temperatures deduced from DSC thermograms. The melting temperatures of the pure PP grades which constitute the matrix are close, and the melting temperature does not change after the addition of talc or glass fiber. However, concerning the crystallization temperature, BA212E presents a relatively high crystallization temperature, while PPH2250 showed a low crystallization temperature, as a consequence after blending this two grades it is supposed to obtain a matrix with a crystallization temperature lower than that of BA212E. The obtained crystallization temperatures for the four composites are approximately close to that of BA212E. Indeed, it is well documented in the literature that talc is an effective nucleating agent that improves the crystallinity of a PP matrix. As a result, we could conclude that the four developed composites retain the good crystallinity of BA212E.

**Table 3.** Meting and crystallization temperatures of the different materials.

|            | $T_c$ (°C) | $T_m$ (°C) |
|------------|------------|------------|
| BA212E     | 126.2      | 165.4      |
| PPH2250    | 112.0      | 163.3      |
| PP-MT30 (A)| 127.1      | 161.6      |
| PP-MT30 (B)| 127.6      | 162.2      |
| PP-MT50    | 127.9      | 161.6      |
| PP-GF30    | 124.7      | 162.2      |

### 3.4. Tensile Fracture and Morphological Analysis

Tensile engineering stress-strain curves of PP-MT30 (A), PP-MT30 (B), PP-MT50, and PP-GF30 are shown in **Figure 11**. For 30 wt.% filler loading, talc-filled PP composite showed a ductile behavior in comparison with GF-filled PP which exhibited a brittle fracture. This may be attributed to the brittle nature of glass fiber compared to talc. Increasing talc content to 50 wt.% decreased sharply the ductility of the material and a brittle fracture was observed. At higher filler loading talc particles tend to agglomerate, they indeed function as stress concentrators, weakening the matrix and lowering the elongation at fracture.

At lower strains, all of the composites display linear behavior, in which the matrix and fibers behave linearly, and nonlinear one at higher strains, which continues until the composite fails completely.

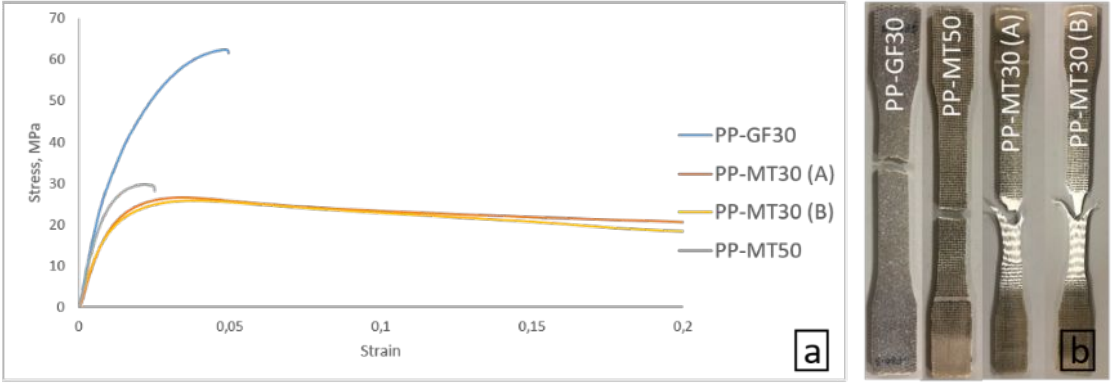

**Figure 11. (a)** Stress-strain curves of the composite materials and **(b)** the corresponding fractured specimens.

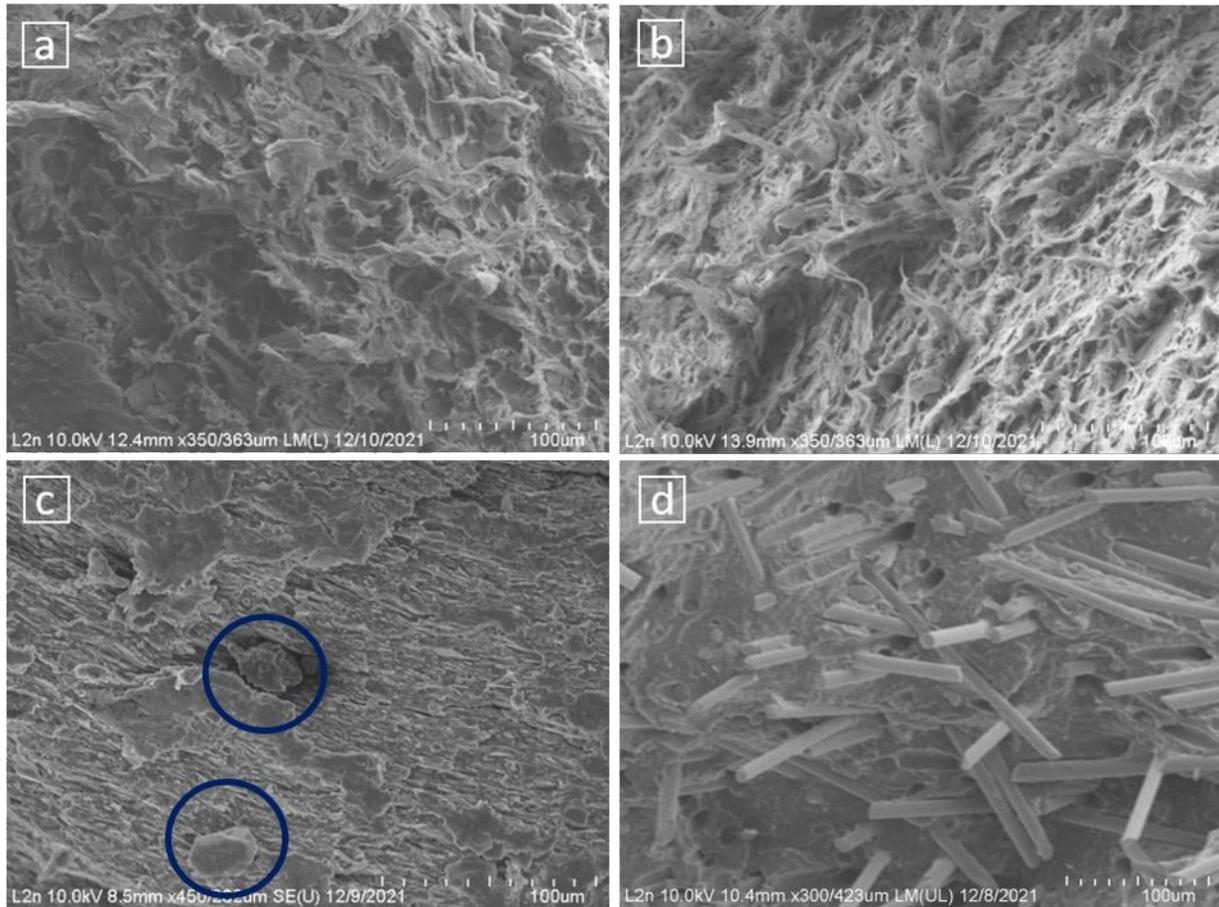

**Figure 12.** SEM micrographs for **(a)** PP-MT30 (A), **(b)** PP-MT30 (B), **(c)** PP-MT50, and **(d)** PP-GF30.

**Figure 12** shows Scanning Electron Microscopy (SEM) images of PP-MT30 (A), PP-MT30 (B), PP-MT50, and PP-GF30 obtained from the fracture surfaces of the tensile specimens. These micrographs confirm the analysis of the stress-strain curves. **Figure 12 (a)** and **(b)** show that talc filler particles are deeply embedded in the fracture surface, indicating good interactions between the filler and matrix; filler dispersion is good, and there are no large visible de-cohesions between the particles and the matrix. However, at higher filler loading, as the case of PP-MT50, talc starts to agglomerate (as seen in **figure 12 (c)**). As a result, the filler-matrix interaction will be reduced and the matrix continuity is replaced by particle–particle contact, resulting in decreased mechanical coherence. [15]

Good mechanical properties were obtained for PP-GF30 due to the high amount of glass fibers and the better interface bonding between glass fiber and PP matrix. It is seen from the SEM micrographs **(Figure 12 (d))** that less glass fibers are pulled out and are broken; similar results were obtained by Soy et al. [30] The presence of a strong bond interface between the glass fiber and the PP matrix is evident in this case.

## 4. Conclusion

Polypropylene has gradually gained a greater place as a pipe material for non-pressure sewerage applications, owing to several advantages over other materials. The Ring Stiffness is the primary property that distinguishes a non-pressure pipe; it is directly proportional to the stiffness of the material. The principal goal of this study is to enhance the stiffness of BA212E; a well-known PP block copolymer used for gravity sewerage applications, in order to obtain higher ring stiffness values, permitting the production of larger pipes while decreasing the wall thickness and therefore improving pipe competitiveness. This was achieved through the incorporation of inorganic fillers, talc and glass fiber. In this study, four different composite materials, PP-MT30 (A), PP-MT30 (B), PP-MT50, and PP-GF30, were characterized in order to be used as a middle layer in a three-layer sewage pipe. Various tensile and flexural tests were performed on specimens produced by injection molding. The four manufactured composites showed a good enhancement of the tensile and flexural moduli. Furthermore, DSC and TGA revealed that the four developed composites retain the good crystallinity the excellent thermal stability of BA212E. According to the MFI test and as it is predicted to use these materials as a middle layer in a three-layer sewage pipe, where BA212E will be used as the outer and inner layer material, PP-MT30 (B) and PP-GF30 were found to be the best candidates for the extrusion process. Also, SEM images of these two materials showed a good filler-matrix interaction in addition to a good filler distribution and dispersion. However, GF-filled PP composite is found to be stronger and stiffer than talc -filled PP composites.

**Acknowledgements**

The authors would like to thank the funders: The French Ministry of Higher Education and the Doctoral School in Sciences and Technology at the Lebanese University (Réseau UT/INSA-UL).

Received: ((will be filled in by the editorial staff))

Revised: ((will be filled in by the editorial staff))

Published online: ((will be filled in by the editorial staff))